\documentclass{JUSTC}

\usepackage{algorithm,algorithmicx,algpseudocode}%
\def\gtap{\ \raise.3ex\hbox{$>$\kern-.75em\lower1ex\hbox{$\sim$}}\ }

\type{Research Articles}
\title{Coupled-channel analysis for vector charmonia and their nature}

\author{Satoshi X. Nakamura}

\email{satoshi@sdu.edu.cn}

\runningtitle{Coupled-channel analysis for vector charmonia\qquad}
\runningauthor{\qquad\qquad S.X. Nakamura}

\address{Institute of Frontier and Interdisciplinary Science, Shandong
University, Qingdao, Shandong 266237, China}

\abstract{
High-precision $e^+e^-\to c\bar{c}$ data (20 final states)
from the BESIII and Belle in $\sqrt{s}=3.75-4.7$~GeV
are analyzed with a semi three-body unitary
coupled-channel model.
Vector charmonium poles are extracted from the 
amplitudes obtained from the fit.
We find well-known $\psi$ states listed in the PDG, and also
several states near open-charm thresholds.
The compositeness of the near-threshold poles suggests 
that 
$\psi(4040)$ could mainly consist of a $D^*\bar{D}^*$-molecule component,
rather than a conventionally accepted quark-model $\psi(3S)$ state.
Also, 
$\psi(4230)$ and $\psi(4360)$ might be substantial mixtures of 
$D_1(2420)\bar{D}$, 
$D_1(2420)\bar{D}^*$, 
$D_s^*\bar{D}_s^*$, and $c\bar{c}$ components.
}


\begin{document}

\maketitle 

\section{Introduction}

Recent BESIII measurements have accumulated
high-quality $e^+e^-$ annihilation cross sections
for various final states covering wide energies.
The measurements showed process-dependent lineshapes associated with
exotic hadron candidates $Y(4230)$ and $Y(4360)$, and discovered
charged charmonium-like $Z_c(3900)$ and $Z_c(4020)$.
The process-dependent lineshapes invalidate the single-channel analysis 
(usual experimental analysis) to determine the vector charmonium
properties such as the mass and width. 
We should analyze the data of different final states simultaneously with
a unified coupled-channel model that respects three-body unitarity. 
Such a coupled-channel analysis is the purpose of this work. 
The resonance properties obtained from the analysis set a primary basis
to study the nature of exotic $Y$ states.
Furthermore, they are a prerequisite to explain how
the process-dependent $Y$ lineshapes come about.
Also, for well-established charmonia [$\psi(4040)$, $\psi(4160)$, $\psi(4415)$],
the coupled-channel analysis will provide new information because
their properties were previously from a simple Breit-Wigner (BW) fit to
the inclusive ($e^+e^-\to$ hadrons) data.
We perform a coupled-channel analysis of the BESIII and Belle data
over $\sqrt{s}=3.75-4.7$~GeV 
for the first time,
and show our fit and
the vector charmonium pole locations.
The full account of the analysis is given in
Ref.~\cite{ours}.

The internal structures of the exotic candidates $Y$ 
have been theoretically studied, and 
$Y(4230)$ [$Y(4360)$] as a $D_1\bar{D}$ [$D_1\bar{D}^*$] hadron molecule
has been proposed~\cite{Ji2022,FZPeng2023}.
Our coupled-channel model has a freedom of generating
the open-charm hadron molecules.
 By examining the compositeness~\cite{sekihara2015},
we explore the hadron-molecule contents in the vector charmonium states.

\section{Coupled-channel model}
\label{sec:model}

\begin{figure}
\begin{center}
\includegraphics[width=.95\textwidth]{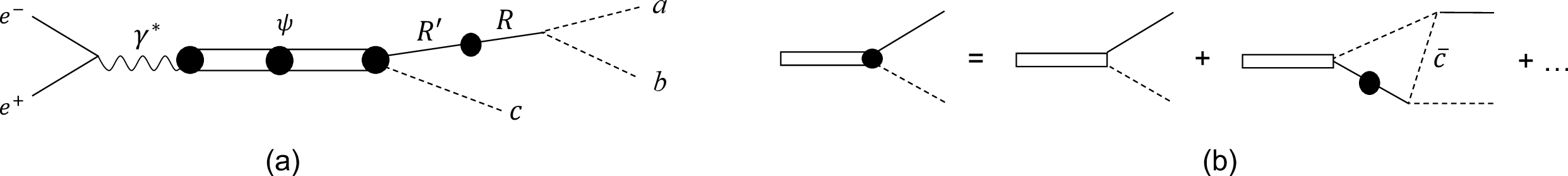}
\end{center}
 \caption{
(a) $e^+e^-\to abc$ mechanism involving 
charmonium excitation. 
(Bare) two-meson resonances $R$ and 
bare charmonium states are represented by 
the solid and double lines, respectively.
Dressed propagators and vertices are indicated by 
the solid circles.
(b) Direct and single triangle decay  mechanisms of charmonium.
 }
\label{fig:diag_1}
\end{figure}

We describe $e^+e^-\to c\bar{c}$ processes
with an approximately three-body unitary coupled-channel model;
see Ref.~\cite{ours} for the full details.
The full amplitude for the three-body ($abc$) final states is given by
\begin{eqnarray}
 A_{abc,e^+e^-} &=& 
\sum^{\rm cyclic}_{abc}
\sum_{RR's_R^z}
\Gamma_{ab,R}\,
\tau_{R,R'}(p_c,E-E_{c})\,
\nonumber\\
&\times&\!\!
\Big[
\sum_{ij}
\bar{\Gamma}^\mu_{R'c,\psi_i}(\bm{p}_c, E)\,
\bar{G}_{ij}(E)\,
\bar\Gamma_{\psi_j,\gamma^*}(E) 
+ \bar\Gamma^\mu_{R'c,\gamma^*}(\bm{p}_c, E) 
\Big]\frac{l_\mu}{s} ,
\label{eq:amp_full_1}
\end{eqnarray}
where the first and second terms in the square bracket are
charmonium-excitation [Fig.~\ref{fig:diag_1}(a)]
and nonresonant (NR) parts, respectively.
Cyclic permutations $(abc)$, $(cab)$, $(bca)$ are indicated by 
$\sum^{\text{cyclic}}_{abc}$;
$E (=\sqrt{s})$ denotes the $abc$ invariant mass and 
$1/s$ is the virtual photon propagator;
the lepton current matrix element is 
$l_\mu (= e\bar{v}_{e^+}\gamma_\mu u_{e^-})$;
$\psi_i$ indicates $i$-th bare $\psi$ state.
A particle $x$'s mass, momentum, and energy are denoted
by $m_x$, $\bm{p}_x$, and $E_x$, respectively.
The symbol $R$ is a two-meson resonance such as $D_1(2420)$,
and $\Gamma_{ab,R}$ is $R\to ab$ vertex.
The amplitude also includes quantities, dressed by quasi-two-body $Rc$
continuum states,
such as
the $Rc$ Green function ($\tau_{R,R'}$),
the $\psi_i\to Rc$ vertex ($\bar \Gamma^\mu_{Rc ,\psi_i}$),
the NR $Rc$ production vertex ($\bar\Gamma^\mu_{Rc,\gamma^*}$),
the $\psi$ production mechanism ($\bar\Gamma_{\psi_i,\gamma^*}$), and
the $\psi$ propagator ($\bar{G}_{ij}$).

\begin{table}[t]
\renewcommand{\arraystretch}{1.0}
\tabcolsep=1.0mm
\caption{\label{tab:Rc}
Quasi two-body ($Rc$) coupled-channels.
}
\begin{tabular}{l} \hline\hline
$D_1(2420)\bar{D}^{(*)}$, $D_1(2430)\bar{D}^{(*)}$,
     $D_2^*(2460)\bar{D}^{(*)}$, $D^{(*)}\bar{D}^{(*)}$, $D_{s1}(2536)\bar{D}_s$
$D_s^{(*)}\bar{D}_s^{(*)}$  \\
$J/\psi\eta$, $J/\psi\eta'$, $\omega\chi_{c0}$, $\Lambda_c\bar{\Lambda}_c$,
$D_0^*(2300)\bar{D}^*$, $f_0J/\psi$, $f_2J/\psi$, $f_0\psi'$,
     $f_0 h_c$, $Z_c\pi$, $Z_{cs}\bar{K}$ \\
\hline\hline
\end{tabular}
\end{table}

We consider $Rc$ channels summarized in Table~\ref{tab:Rc}.
For unstable charmed mesons other than $D^*_0(2300)$,
we use a BW form for $\tau_{R,R'}$,
which causes the partial three-body unitarity violation.
$D_0^*(2300)$, $f_{0(2)}$, and $Z_{c(s)}$ are poles in
$L=0$ $D\pi$,
$L=0 (2)$ $\pi\pi-K\bar{K}$, and
$J^{PC}=1^{+-}$ 
$D^*\bar{D}-D^*\bar{D}^*-J/\psi\pi-\psi'\pi-h_c\pi-\eta_c\rho$
($J^{P}=1^{+}$ $D_s^*\bar{D}-D_s\bar{D}^*-J/\psi K$)
coupled-channel scattering amplitudes, respectively.

The nonperturbative coupled-channel
$Rc$ scattering is driven by
bare $\psi$-excitations,
(on-shell) particle-exchanges, and
short-range contact interactions,
in a way to satisfy the three-body unitarity.
The bare $\psi$ states are then dressed by the $Rc$ continuum states to form charmonium resonance states.
Besides, the short-range contact interactions alone may 
generate hadron-molecule states.

\section{Fit results}
\label{sec:fit}

We fitted $e^+e^-\to c\bar{c}$ cross-section data (20 final states) 
as well as available invariant-mass distribution data. 
Five bare $\psi$ states were needed for reaching a
reasonable fit. 
Furthermore,
we included $\psi(4660)$ and $\psi(4710)$ BW amplitudes to fit data for 
$\sqrt{s}\gtap 4.6$~GeV.
With 200 fitting parameters, our default fit reached
$\chi^2/{\rm ndf}=2320/(1635-200)\simeq 1.6$.

\begin{figure}
\includegraphics[width=1.\textwidth]{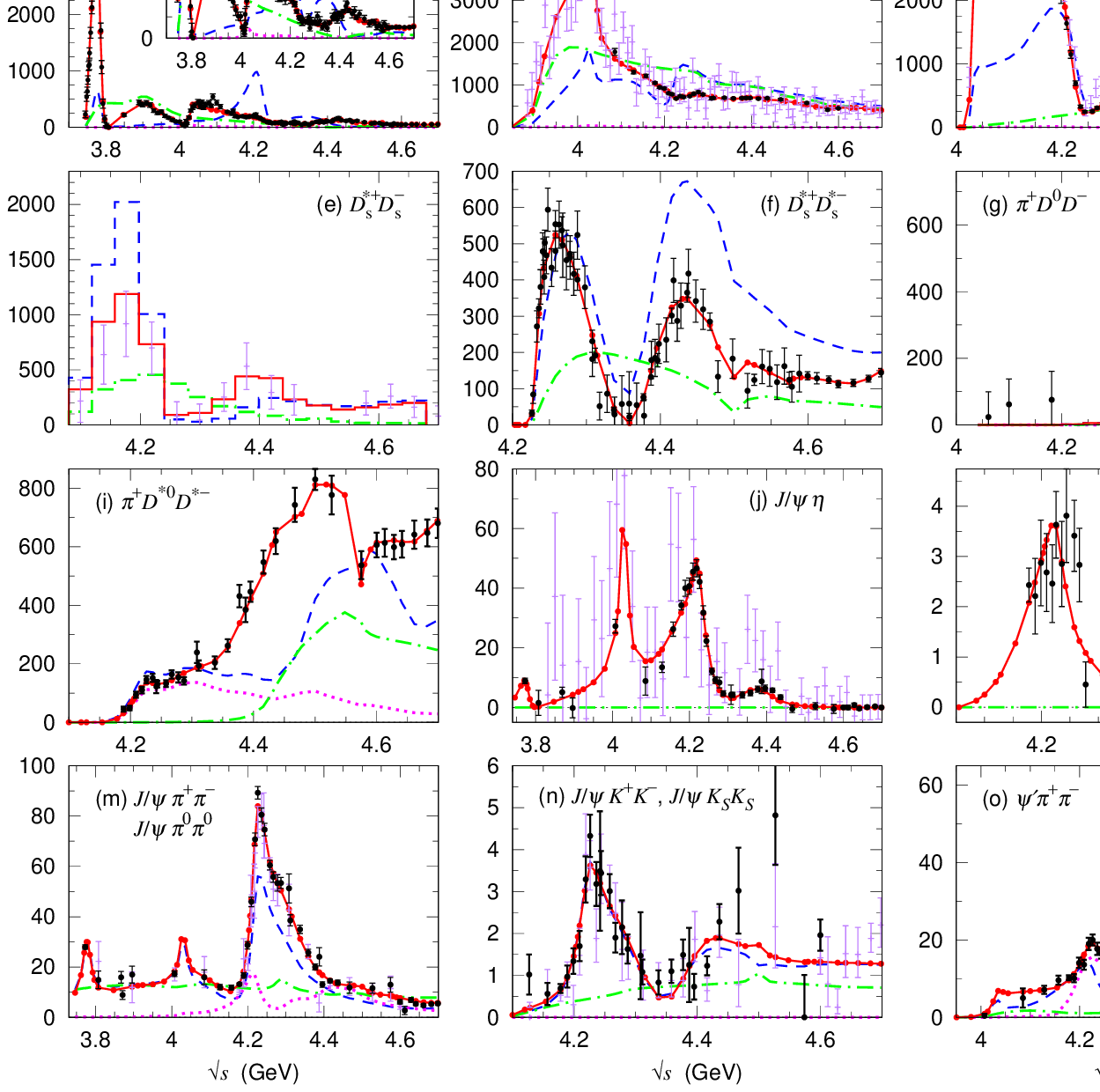}
 \caption{\label{fig:xs-all5}
$e^+e^-$ annihilation
cross sections (unit:pb); each panel indicates the final state;
$\sqrt{s}$ is the total energy. 
Full calculations are shown by 
the red points, connected by lines.
The direct decays, single-triangle, and nonresonant contributions
are shown by 
the blue dashed, magenta dotted, and green dash-dotted curves, respectively.
Figures taken from Ref.~\cite{ours} where references for the data are given. 
 }
\end{figure}
Some of the fit results are shown in Fig.~\ref{fig:xs-all5}.
For more fit results, see Ref.~\cite{ours}.
The default-fit results are shown 
by the red circles connected by lines.
Also, contributions from 
the direct-decay and single-triangle rescattering mechanisms
defined in Fig.~\ref{fig:diag_1}(b) are shown by the blue-dashed and
magenta dotted curves.

Remarks are in order:
(i) The data show several cusp structures at thresholds such as:
$D_1(2420)\bar{D}^*$ (4431~MeV) in Fig.~\ref{fig:xs-all5}(a);
$D_1(2420)\bar{D}$ (4289~MeV) in Fig.~\ref{fig:xs-all5}(b);
$\Lambda_c\bar{\Lambda}_c$ (4573~MeV) in Fig.~\ref{fig:xs-all5}(i).
Our model generates these cusp structures with
the short-range contact interactions.
(ii) The short-range interactions generate hadron-molecule poles that cause significant threshold enhancements, 
as indicated by differences between the red solid and blue dashed curves in 
Figs.~\ref{fig:xs-all5}(a)-(f).
\begin{figure}
\includegraphics[width=0.99\textwidth]{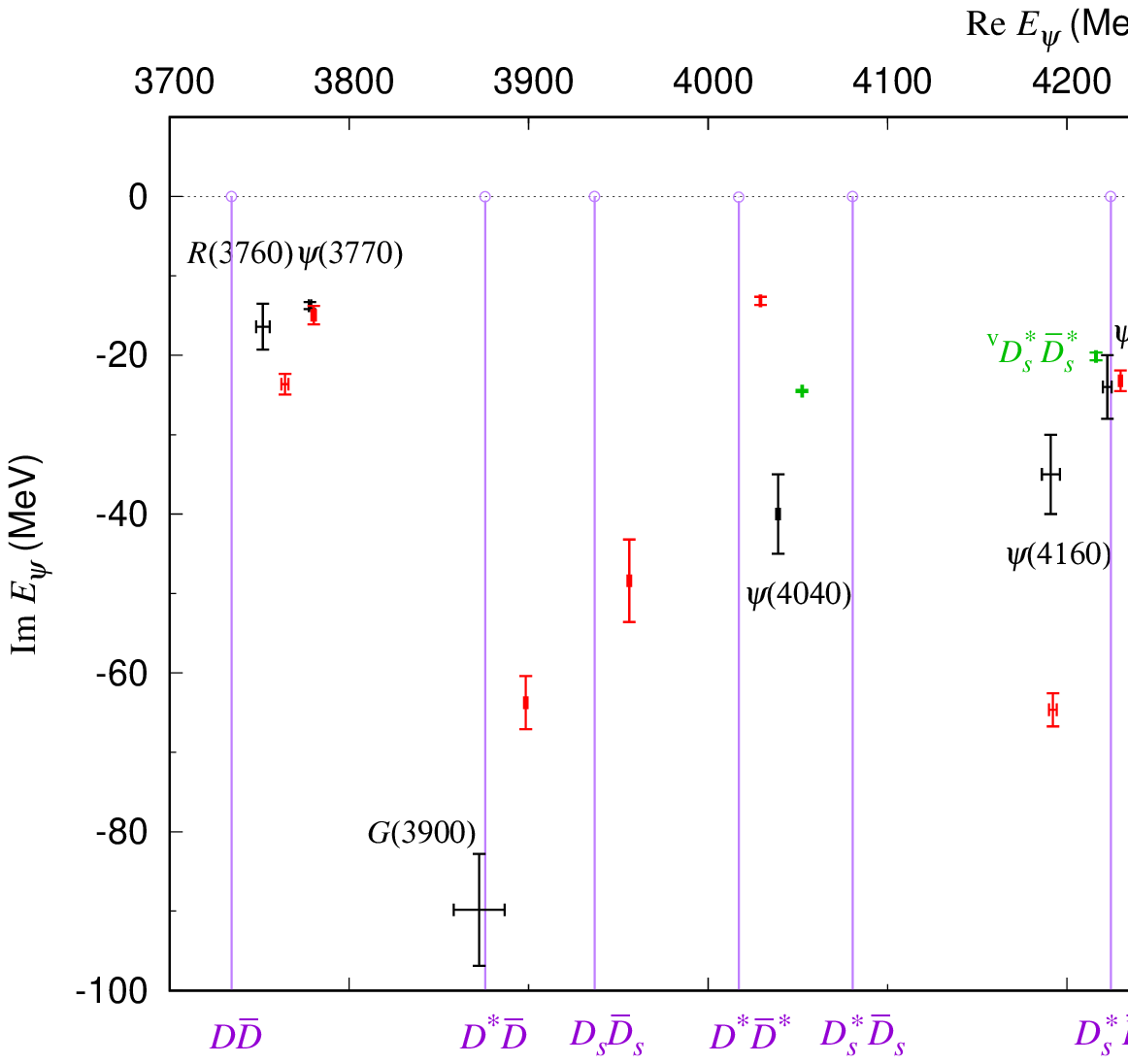}
  \caption{Vector charmonium poles ($E_\psi$) and their uncertainties. 
Red, blue, and green points indicate pole locations of
resonances (located on unphysical sheets of open channels), 
bound, and virtual states, respectively;
the bound (virtual) states are on 
the physical (unphysical) sheets of the nearest-threshold channels, respectively. 
Black points indicate
$\psi$ states listed in PDG~\cite{pdg}, 
${\cal R}(3760)$~\cite{bes3_r3780}, 
$G(3900)$~\cite{bes3_DD}, and 
$Y(4320)$~\cite{bes3_jpsi-pippim}.
Branch points (thresholds) and cuts for 
open-charm channels indicated at the bottom 
are shown by 
open circles and accompanying vertical lines, respectively.
Figure taken from Ref.~\cite{ours}.
 }
\label{fig:cc}
\end{figure}
(iii) As a consequence of the coupled-channel fit,
our model creates 
common structures in different processes,
even when not necessarily required by the data.
For example, $\psi(4040)$ peaks appear in 
$D^*\bar{D}$ [Fig.~\ref{fig:xs-all5}(b)] and 
$D_s\bar{D}_s$ [Fig.~\ref{fig:xs-all5}(d)] to fit the data,
and they also appear in others [Fig.~\ref{fig:xs-all5}(j,m,o,p)]
for which data are lacking at the peak.
(iv) The $J/\psi K^+K^-$ data [Fig.~\ref{fig:xs-all5}(n)] show
an enhancement suggesting $Y(4500)$~\cite{bes3_jpsi-kpkm}.
However, our model does not fit it since
the data is rather fluctuating in this region, and 
the $J/\psi K_SK_S$ data does not indicate the same enhancement.

\section{Vector-charmonium poles and compositeness}
\label{sec:pole}

The coupled-channel amplitude obtained from the fit
is analytically continued for 
searching vector charmonium poles $E_\psi$ on
unphysical sheets of the open channels and 
physical sheets of the closed channels, or slightly deviated from this condition.
The range is 
$3.75 < M < 4.7$~GeV ($M\equiv {\rm Re} [E_{\psi}]$),
and $\Gamma\equiv -2\times{\rm Im} [E_\psi] < 0.2$~GeV.

Fourteen states are found, as shown in Fig.~\ref{fig:cc} along with 
experimental analysis results.
Compared to the uncertainties estimated in 
the experimental single-channel analyses,
our pole uncertainties are generally smaller. 
This is probably because our pole values are constrained by 
the data of the various processes and some of the data are very precise. 
Our result includes all the counterparts of 
the vector charmonium states 
($M>3.75$~GeV) listed in the PDG~\cite{pdg}, 
although sizable differences can be seen
such as the $\psi(4040)$ width and the $\psi(4415)$ mass and width. 
Actually, 
the $\psi(4040)$, $\psi(4160)$, and $\psi(4415)$
resonance parameters in the PDG are from a simple BW fit to
the $R$ values~\cite{bes2-R}.
Such a simple analysis could have caused 
artifacts in the resonance parameters. 
More states are found near the open-charm thresholds.
Those located near the $D_{s}^{(*)}\bar{D}_s^{(*)}$ and
 $D_{s1}\bar{D}_s$ thresholds 
are found for the first time in the present analysis.

We calculate the
compositeness~\cite{sekihara2015} of the poles 
as a qualitative measure of 
the internal structure.
A continuum (two-body)
$Rc$ channel content in a state is denoted by 
$X_{Rc}$ (compositeness) and, combined with 
an elementariness $Z_a$ of a bare state $a$, 
it satisfies $\sum_{Rc} X_{Rc} + \sum_a Z_a=1$.
However, it is noted that the compositeness should be viewed with caution
because $X_{Rc}$ is generally complex and 
its imaginary part and negative real part are
difficult to interpret. 
Also, the $X_{Rc}$ also depends on the choice of the form factors.

\begin{table}
\begin{center}
\renewcommand{\arraystretch}{0.9}
\caption{\label{tab:comp1} 
Compositeness $X_{Rc}$
of some of vector charmonium states shown in 
Fig.~\ref{fig:cc}.
Hyphens indicate $|X_{Rc}|<0.01$.
Table taken from Ref.~\cite{ours}.
}    
\begin{tabular}{lcccc}\hline\hline
$Rc\ \backslash\ \psi$
& 
$\psi(4040)$                         & 
${}^{\rm v}D_{s}^*\bar{D}^*_s$	     &
$\psi(4230)$                         &
$\psi(4360)$      		     \\\hline
$D\bar{D}$ &                --&                  --&--&--\\
$D^*\bar{D}$ &	       $-0.01+ 0.01i$&	         --&--&--\\
$D^*\bar{D}^*$ &	       $ 0.86+ 0.22i$&	         $ 0.01+ 0.02i$&--&--\\
${D}_1\bar{D}$ & 	       --& 	         $ 0.23-0.12i$&$ 0.18+ 0.13i$&$ 0.31-0.06i$\\
${D}_1\bar{D}^*$ & 	       --& 	         $ 0.04-0.06i$&$ 0.09+ 0.05i$&$ 0.29-0.16i$\\
${D}_2^*\bar{D}^*$&	       --&	         $ 0.02-0.01i$&--&--\\
$D_s^*\bar{D}_s^*$ &	       --&	         $ 0.50+ 0.25i$&$ 0.35-0.27i$&--\\
Sum&			       $ 0.86+ 0.19i$&			         $ 0.83+ 0.12i$&$ 0.65-0.13i$&$ 0.61-0.27i$\\
\hline\hline
  \end{tabular}
\end{center}
\end{table}

In Tables~\ref{tab:comp1},
we present the compositeness for some of the vector 
charmonium states shown in Fig.~\ref{fig:cc}.
Some states have large compositeness.
It is interesting to find that the well-established
$\psi(4040)$ has $X_{D^*\bar{D}^*}=0.86$.
Conventionally, 
this state has been assigned to the quark-model $\psi(3S)$ state 
as an input to determine the quark-model parameters.
Our result 
might be casting a doubt to this conventional practice.
In literature~\cite{Ji2022,FZPeng2023}, it has been argued that
$\psi(4230)$ and $\psi(4360)$ can be
$D_1(2420)\bar{D}$ and $D_1(2420)\bar{D}^*$ molecules, respectively.
However, our result suggests more complex structures.
Due to the coupled-channel dynamics, 
molecular states 
[$D_1(2420)\bar{D}$, $D_1(2420)\bar{D}^*$, and $D_s^*\bar{D}^*_s$] 
and also bare $\psi$ states are substantially mixed
in
$\psi(4230)$, $\psi(4360)$, and 
${}^{\rm v}D_{s}^*\bar{D}^*_s$ states.

\section{Summary}
\label{sec:summary}
A global coupled-channel analysis of most of the available
$e^+e^-\to c\bar{c}$ data (20 final states) in $\sqrt{s}=3.75-4.7$~GeV
was performed.
Reasonable fits were obtained for both 
cross-section and invariant-mass distribution data.
We extracted vector-charmonium poles including
not only familiar vector charmonia, but
also those near the open-charm thresholds.
The compositeness of the states suggested that open-charm
hadron-molecular structures dominated in many states.
The $\psi(4230)$ and $\psi(4360)$ states are not simple 
$D_1\bar{D}$ and $D_1\bar{D}^*$ molecules, respectively,
but are the mixtures of $c\bar{c}$ and $D_1\bar{D}$, $D_1\bar{D}^*$, and $D_s^*\bar{D}_s^*$ molecules.
A large $D^*\bar{D}^*$ compositeness in $\psi(4040)$ was also suggested.
%




\end{document}